\documentclass[11pt]{article}
\usepackage[dvips]{graphics}
\begin{document}
\input epsf.sty

%%%%%%%%%%%%%% %%% Definitions:  %%%%%%%%%%%%%% 
%definition de la font pour
%R,C,N 
% \font\blackboard=msbm10 % scaled \magstep1 % \font\blackboards=msbm7
\font\blackboardss=msbm5 
% \newfam\black \textfont\black=\blackboard 
%
%\scriptfont\black=\blackboards \scriptscriptfont\black=\blackboardss 
%
%\def\Bbb#1{{\fam\black\relax#1}} %definition de la font pour l'identite
%\font\ninerm=cmr9 %\def\uniset{\rlap{\ninerm 1}\kern.15em 1}
%\def\e{\mathop{\rm e}\nolimits} 
%Fractions 
\def\half{\scriptstyle{\frac{1}{2}}}
\def\halft{\textstyle{\frac{1}{2}}} 
\def\osqrt{\textstyle{\frac{1}{\sqrt2}}}
\def\lsqrt{\textstyle{\frac{\l}{\sqrt2}}} 
\def\phalf{\textstyle{\frac{\pi}{2}}}
\newcommand{\fscr}[2]{\scriptstyle \frac{#1}{#2}} 
%Definitions 
\def\cc{\hbox{C\kern-0.55em\raise0.4ex\hbox{$\scriptstyle |$}}}
\def\zz{\hbox{{\sf Z}\kern-0.45em\raise0.0ex\hbox{\sf Z}}}     
\def\nat{\hbox{{\sf |}\kern-0.45em\raise0.0ex\hbox{\sf Í}}}     
\def\rr{\hbox{R\kern-0.55em\raise0.4ex\hbox{$\scriptstyle |$}}}  
\def\ena{1\hskip-0.25truecm 1}  
\def\zita{{\zz}^{ 2}_{N}}
\def\zitas{{\zz}^{* 2}_{N}}
\def\zi{{\zz}^{2r+1}_{2}}
\def\zist{{\zz}^{* 2r+1}_{2}}
\def\ccd{{\cc\hskip0.2truecm}^{2}}
\def\cct{{\cc\hskip0.2truecm}^{3}}                                          

\def\pf{{\it pd \/}}
\def\ie{{\it i.e \/}}
\def\tr{{\rm Tr \/}}
\def\eg{{\it e.g \/}} 
\def\cf{{\it c.f \/}} 
\def\viz{{\it viz. \/}}
%Decorated alphabet 
\def\ad{a^\dagger} 
\def\ab{\bar{\alpha}} 
\def\nub{\bar{\nu}}
\def\hi{\chi_{klm}} 
\def\udp{U_{\lambda}^{\dagger}}
\def\udm{U_{-\lambda}^{\dagger}} 
\def\utp{\tilde{U}_{\lambda}}
\def\utm{\tilde{U}_{-\lambda}} 
\def\up{U_{\lambda}} 
\def\um{U_{-\lambda}}
%Sigmas 
\def\so{\sigma_{1}} 
\def\st{\sigma_{2}} 
\def\sth{\sigma_{3}}
\def\sp{\sigma^{+}} 
\def\sm{\sigma^{-}} 
%Abbreviations
\def\ovl{\overline}

%
%Math I,R,C %\def\IR{\rm I}\!{\rm R}
%\newcommand{\dbl}[2]{\rm#1\hskip-.5em \rm#2} %Bra-kets, absolutes
\newcommand{\bra}[1]{\left<#1\right|} \newcommand{\ket}[1]{\left|#1\right>}
\newcommand{\braket}[1]{\left<#1\right>}
\newcommand{\inner}[2]{\left<#1|#2\right>}
\newcommand{\sand}[3]{\left<#1|#2|#3\right>}
\newcommand{\proj}[2]{\left|#1\left>\right<#2\right|} %
\newcommand{\rbra}[1]{\left(#1\right|} \newcommand{\rket}[1]{\left|#1\right)}
\newcommand{\rbraket}[1]{\left(#1\right)}
\newcommand{\rinner}[2]{\left(#1|#2\right)}
\newcommand{\rsand}[3]{\left(#1|#2|#3\right)}
\newcommand{\rproj}[2]{\left|#1\left)\right(#2\right|}
\newcommand{\absqr}[1]{{\left|#1\right|}^2}
\newcommand{\abs}[1]{\left|#1\right|}
 %Derivatives
\newcommand{\pl}[2]{\partial_{#1}^{#2}} 
\newcommand{\plz}[1]{\partial_{z}^{#1}}
\newcommand{\plzb}[1]{\partial_{\overline{z}}^{#1}} 
\newcommand{\zib}[1]{{\overline{z}}^{#1}} 
%Matrices
\newcommand{\mat}[4]{\left(\begin{array}{cc} #1 & #2 \\ #3 & #4
\end{array}\right)} % 
\newcommand{\col}[2]{\left( \begin{array}{c} #1 \\ #2
\end{array} \right)} 
%Gr-alphabet 
\def\a{\alpha} 
\def\b{\beta} 
\def\g{\gamma}
\def\d{\delta} 
\def\e{\epsilon} 
\def\z{\zeta} 
\def\th{\theta} 
\def\f{\phi}
\def\la{\lambda} 
\def\m{\mu} 
\def\p{\pi} 
\def\om{\omega}  
\def\D{\Delta} 
\def\zb{\bar{z}} 
%Others 
\newcommand{\lag}[2]{L_{#1}^{#2}(4\l^{2})}
\newcommand{\mes}[1]{d\mu(#1)} 
%Command abbreviations 
\def\nd{\noindent}
\def\nn{\nonumber} 
\def\cap{\caption} 
\def\cline{\centerline}
\newcommand{\be}{\begin{equation}} 
\newcommand{\ee}{\end{equation}}
\newcommand{\ba}{\begin{array}} 
\newcommand{\ea}{\end{array}}
\newcommand{\bea}{\begin{eqnarray}} 
\newcommand{\eea}{\end{eqnarray}}
\newcommand{\beann}{\begin{eqnarray*}} 
\newcommand{\eeann}{\end{eqnarray*}}
\newcommand{\bfg}{\begin{figure}} 
\newcommand{\efg}{\end{figure}}
%%%%%%%%%%%%%%%%%%%%%%%%%%%%%%%%%%%%%%%%%%%%%%%
\def\ucn{U_{\scriptstyle  CN}}
\def\uca{U_{\scriptstyle  f}}

%%%%%%%%%%%%%%%%%%%%%%%%%%%%%%%%%%%%%%%%%%%%%%%%%%%%%%%%%%%%%%%%%%%%%%%%%%%%% %
%\title{Mixed state entanglement of finite quantum systems} % %\author{Demosthenes
%Ellinas\affil{{\tt dimellin\char"40talos.cc.uch.gr}\\ %Department of
%Mathematics, University of Crete \\ %P.O.Box 1470, Heraklion GR-714 09, Crete,
%Greece}}
%\hspace*{10.0cm}\parbox[t]{3.0cm}{FTUV/94-42\\IFIC/94-47\\}  
%Forthcoming in: Journal of Nonlinear Mathematical Physics
%
%
\title{ Quantization of Soliton Cellular Automata \footnote{Submitted to Journal of Nonlinear Mathematical Physics.
 Special Issue of Proccedings of NEEDS'99.}}

%\begin{flushright}
            
%September 1994}
%%\end{flushright}
\author{Demosthenes Ellinas\footnote{\tt ellinas@science.tuc.gr} \ \ 
Elena P. Papadopoulou\footnote{\tt elena@science.tuc.gr} \ \ and
Yiannis G. Saridakis 
\footnote{\tt yiannis@science.tuc.gr}\\
\\
Department of Sciences\\
Section of Mathematics\\
Technical University of Crete\\
GR - 73 100 Chania Crete Greece}

\maketitle
\begin{abstract}
  A method of quantization of classical soliton cellular automata
  (QSCA) is put forward that provides a description of their time evolution
  operator by means of quantum circuits that involve quantum gates from which the associated Hamiltonian
describing a quantum chain model is constructed. The
  intrinsic parallelism of QSCA, a phenomenon first known
  from quantum computers,  is also emphasized.
\end{abstract}

{\it Introduction.} Soliton cellular automata (SCA), is a class of cellular automata\cite{wolf}
 operating on binary sequences with an updating rule function $f$
 for each cell, that depends on past and present time cells, the
 number of which determines the {\it radius r} of the automaton\cite{steig}. Contrary
 to usual CA ( that evolve their cells by means of past time cells
 only ), SCA exhibit a variety of evolution patterns\cite{cca,takh,bso} that is mainly
 known to characterize the temporal behavior of solutions of non
 linear PDE's\cite{soliton}, namely: periodic evolution of {\it particles} (i.e localized
 groups of binary cells ), or {\it solitonic} type of scattering of digital
 particles, or even {\it breathing} modes of oscillations between particles.
 All these properties have motivated a number of suggestive
 applications for a new kind of computational architecture that will
 utilize these evolution patterns of SCA in order to provide a
 "gateless" implementation of logical operations\cite{steig}.
  Towards a physical microscopic realization of these suggestions,
 envisaged in the context of the new paradigm of
 {\it Quantum Computing}, it is plausible to formulate SCA in
 terms of Quantum Mechanics and to investigate the possible
 quantum effects in their time evolution. To this end we put
 forward here the quantization of classical SCA, and point out their quantum parallelism.

{\it Quantization.} Let the product Hilbert space ${\cal H}=\otimes_{i\in \scriptstyle \zz}{\cal H}_{i}$  where   
${\cal H}_i =
span \{ \ket{0}, \ket{1}\}\approx \cc^2$. Take $n\in {\bf N} $, and consider the subspace
$H\subset \cal H$ where
$H = \otimes_{k=-r}^{r}  {\cal H}_{n-k}$, then define the vectors
\be
\ket{a^t}\equiv \otimes_{i=r}^{1}\ket{a_{n-i}^{t+1}}\otimes\ket{a_n^t}\otimes_{j=1}^{r}\ket{a_{n+j}^{t}} \in H \;,
\ee
\nd and the dual vectors $\bra{a^t} \in \tilde{H}$, with 
orthogonormality relation
\bea
\inner{a^t}{b^t}&=& \Pi_{i=r}^{1}\inner{a^{t+1}_{n-i}}{b_{n-i}^{t+1}}
\cdot \inner{a_n^t }{b_n^t }\cdot 
\Pi_{j=1}^{r}\inner{a^{t}_{n+j}}{b_{n+j}^{t}}\nn \\
&=& \mbox{}
\Pi_{i=r}^{1}\d_{a^{t+1}_{n-i},b_{n-i}^{t+1}}
\cdot \d_{a_n^t ,b_n^t }\cdot 
\Pi_{j=1}^{r}\d_{a^{t}_{n+j},b_{n+j}^{t}}\;.
\eea

\nd Recall that for $(x,y)\in \zz_2^2 $ the Kronecker 
delta function is defined as $\d_{x,y}= 1\oplus x \oplus y $,
where $\oplus$ denotes $\tt XOR$, i.e modulo-$2$ addition. Now take
$\zist \equiv \zi \setminus O^{2r+1}$, where $O^{2r+1}$ is the
$2r+1$-fold null string,
$\zz^{{\dagger} 2r+1}_2 \equiv \zz^{2r+1}_2 \setminus \{a_0 \}$
where $f(\{a_0 \})=O^{2r+1}$, the
preimage of the null string, and define
the one-to-one function\cite{takh} $f:\zist \rightarrow \zz^{{\dagger} 2r+1}_2$, as
$f(a^t )= a^{t+1}$, where 
$a^t \equiv \{ a_{n-r}^{t+1},\ldots,a_{n-1}^{t+1},
a_{n}^{t},a_{n+1}^{t}
,\ldots,a_{n+r}^{t}\} $, and
$a^{t+1} \equiv \{ a_{n-r}^{t+1},\ldots,a_{n-1}^{t+1},
a_{n}^{t+1},a_{n+1}^{t}
,\ldots,a_{n+r}^{t}\} $, where the updated bit takes the value
$a_{n}^{t+1} = 1\oplus_{i=r}^{1} a_{n-i}^{t+1} \oplus_{j=0}^{r}
 a_{n+j}^{t} $.

Introduce now the quantization of classical cellular automaton by means of the following quantization diagram.
\vspace{0.3cm}
\[
\begin{array}{rcccl}
 & \zist & \stackrel{f}{\longrightarrow} & 
 \zz^{{\dagger} 2r+1}_2 & \\
 \rho & \downarrow & & 
\downarrow & 
\rho \\
 & H & 
 \begin{array}[t]{c}\longrightarrow \\ \uca \end{array}& 
H & 
\end{array} 
\]
%\vspace{0.2cm}

\nd This diagram implies that:
\be
\uca \circ \rho (a^t )=\uca\ket{a^t }=\ket{f(a^t )}=\ket{a^{t+1}}= 
%\nn \\
\rho \circ f(a^t )= \rho (a^{t+1})=\ket{a^{t+1}}\;.
\ee

\nd The transition operator $U_f $ implements in the space of qubits $H$ the
update rule $f$ of the classical bits of the automaton, and acts trivially as the unit operator in the rest space. Since in order to establish
the one-to-one property for $f$ we have excluded the null string from its domain of values
the associated operator $U_f \in {\it End}H$, is a partial isometry in $H_0^\bot$,
the 
orthogonal complement space of $H_0 \equiv {\it span}\{ \ket{0}\}$, assuming the
decomposition $H=H_0 + H_0^\bot$. To be more specific, let
$\uca=\sum_{a^t \in \zist}\proj{f(a^t )}{a^t }$, then its isometric property is due to
the following relations:
\bea
\uca {\uca}^\dagger &=&\sum_{(a^t , b^t )\in \zist}\ket{f(a^t )}\inner{a^t }{b^t }
\bra{f(b^t )} =\sum_{a^t \in \zi \setminus \{a^t_0 \} }\proj{a^t }{a^t }\;,\nn \\
%\nd {\rm and}\nn \\
{\uca}^\dagger \uca &=& \sum_{(a^t , b^t )\in \zist}\ket{b^t }\inner{f(b^t ) }{f(a^t )}
\bra{a^t }= \sum_{a^t \in \zi \setminus \{ 0 \} }\proj{a^t }{a^t }\;,
\eea

\nd where the preimage of the null string for e.g the simplest case of $r=2$ is
$a^t_0 = 00100 $. 

Choosing a basis of vectors in $H$, e.g the computational basis
${\cal B}\equiv \{ \ket{x_1 }\otimes\cdots \otimes \ket{x_{2r+1}} : \{x_i \}_{i=1}^{2r+1} \in \zz^{2r+1}_2 \}$,
we can obtain a $2^{2r+1}\times 2^{2r+1}$ matrix representation of $\uca$ viz.
\bea
& &\pi_r (U_f )= \nn \\
& & \sum_{a^t \in \zz^{2r+1} }
\otimes_{i=r}^{1}
\mat{\d_{ 0,a_{n-i}^{t+1} } }{0}{0}{\d_{1,a_{n-i}^{t+1} } }
\otimes 
\mat{\d_{0,a_{n}^{t+1}} \d_{0,a_{n}^{t}}}
{\d_{0,a_{n}^{t+1}} \d_{1,a_{n}^{t}}}
{\d_{1,a_{n}^{t+1}} \d_{0,a_{n}^{t}}}
{\d_{1,a_{n}^{t+1}} \d_{1,a_{n}^{t}}}
\otimes_{j=1}^{r}
\mat{\d_{0,a_{n+j}^{t}} }{0}{0}{\d_{1,a_{n+j}^{t}}}\;.
\eea

\nd More explicitly if we partition the basis $\cal B$ into two orthogonal compliments
corresponding to invariant and non invariant subspaces of $\uca$ we will obtain for e.g the case of radius $r=2$, the transition matrix :
\be
\pi_2 (\uca ) = \mat
{\bf 1}{\bf 0}
{\bf 0} {\sigma_1^{'}} \;,
\ee

\nd where ${\bf 1}$,${\bf 0}$, stand for the square $16$-dimensional unit and null 
matrices respectively, while $\sigma_1^{'}$, is a $N=16$-dimensional matrix with elements along the
 main antidiagonal $(a_{1,N}, a_{2,N-1},\ldots,a_{N,1})=
 (1, \ldots , 1,0 )$, and all others been zero.

%%%%%%%%%%%%%%%%%%%%%%%%%%%%%%%%%%%%%%%%%%%%%%%%%%%%%%%%%%%%%%%%%%%%%%%%%%%%%%%%%%
%\begin{figure}[hptb]
%\begin{center}
%\scalebox{0.2}{\includegraphics{hontro.eps}}
%\end{center}
%\caption{
%Fig. 1. The factorization of $\uca$ in terms of quantum gates for radius $r=2$.\\
%Fig. 2. Quantum circuit implementation of 
%FRT for three quantum BS's with $r=2$.}
%\label{fig12}
%\end{figure}

%\vskip 0.5cm
%\be
%\ucn =\mat
%{{\bf 1}_2 }{{\bf 0}_2}
%{{\bf 0}_2 }{\sigma_1 }
%\ee
%%%%%%%%%%%%%%%%%%%%%%%%%%%%%%%%%%%%%%
\vspace{1.5cm}
%\begin{figure}[h]
%\epsfysize=17.5cm
%\centerline{\epsffile{control.eps}}
%\vspace*{-14.5cm}
%\caption{  The factorization of $\uca$ in terms of quantum gates for radius
%$r=2$.\\
Fig. 1. The factorization of $\uca$ in terms of quantum gates for radius
$r=2$.\\
Fig. 2. Quantum circuit implementation of 
FRT for three quantum BS's with $r=2$.
%\end{figure}

\vskip 0.5cm

{\it Hamiltonian model.} In order to construct a Hamiltonian model associated with the QSCA that generates the total evolution of the automaton, and in view of the fact that the time step evolution operator is determined by unitary $\tt NOT$ viz. $U_N^i \ket{i}=\ket{1\oplus i}$ and
 $\tt CONTROL-NOT(CN)$ viz. $U_{CN}^{ij}\ket{ij}=\ket{i i\oplus j}$
 gate operators, cf.
 its explicit factorization in Fig. 1,
we recall here first the Hermitian operators corresponding to those gates
[$ \ket{i}\otimes\ket{j}$ is abbreviated to
$\ket{ij}$].
The quantum negation of
the $i$-th qubit is given by the matrix $U_N^i = \mat{0}{1}{1}{0} _i$ acting non trivially
only in the $i$-th subspace, with associated Hermitian generator given by
$H_N^i = \frac{1}{2}(\sigma_z^i + \sigma_x^i )$, in terms of Pauli sigma matrices\cite{qm}.
For the conditional negation of $\tt CN$ gate with $i$-control and $j$-target qubits and
$i< j$
the unitary matrix $U_{CN}^{ij}=\proj{0}{0}\otimes{\bf 1}+\proj{1}{1}\otimes\sigma_x =
\exp(i\pi H_{CN}^{ij})$ is generated by the Hamiltonian $H_{CN}^{ij}=\frac{1}{2}
({\bf 1}-\sigma^i _z )(\sigma_x^j -{\bf 1})$. Correspondingly for $i> j$, 
$U_{CN}^{ij}={\bf 1}\otimes \proj{0}{0} + \sigma_x \otimes \proj{1}{1} =
\exp(i\pi H_{CN}^{ij})$, and $H_{CN}^{ij}=\frac{1}{2}
({\bf 1}-\sigma^j _z )(\sigma_x^i -{\bf 1})$. These gate operators can be utilized to factorize
the $\uca$ that is used to update each qubit in a given string of qubits of a QSCA.
To this end in Fig.1 we construct the quantum circuit that provides such a factorization
for the special case $r=2$; the generalization to any $r$ is straightforward. Specifically
the input state vector in Fig. 1 reads $\ket{a^t}\equiv \ket{a^{t+1}_{n-2},a^{t+1}_{n-1},
a^{t}_{n},a^{t}_{n+1},a^{t}_{n+2}}$, and is tranformed by $\uca$ to 
$\ket{a^{t+1}}\equiv \uca\ket{a^t}=\ket{a^{t+1}_{n-2},a^{t+1}_{n-1},
a^{t+1}_{n},a^{t}_{n+1},a^{t}_{n+2}}=U_N^n U_{CN}^{n+2,n}U^{n+1,n}_{CN}U^{n-1,n}_{CN}U^{n-2,n}_{CN}\ket{a^t_n}$, where $\ket{a^{t+1}_n }=\ket{1\oplus a_{n-2}^{t+1} \oplus a_{n-1}^{t+1}
\oplus
 a_{n}^{t}\oplus
a_{n+1}^{t}\oplus
a_{n+2}^{t}} $. In Fig. 1 graphically the $\tt CN$ gate is indicated as a vertical 
line segment with a bullet (control qubit) and a crossed circle (target qubit) at its ends
\cite{bar}.
Then the total unitary evolution operator $U_f = \prod_i ^\infty U_i ^f \equiv \ldots U_2 ^f 
U_1 ^f U_0 ^f $, is the product of update operators for each qubit in the automaton starting
with the first one $U_0 ^f$, from which all later ones are determined by shifting i.e
$U_{i+1}^f={\bf 1}\otimes U_i ^f $. The total Hamiltonian $H_f =\lim_{t\rightarrow 0}
\frac{1}{i}\frac{dU_f }{dt}=\sum_{i\geq 0}H_i ^f$, is the sum of all Hamiltonians
generationg each evolution operator $\{U_i ^f ,  i\geq 0\}$. Since $U_i ^f = U_N ^i \prod_{k=1}^{r}
U_{CN}^{i+k , i}\prod_{k=1}^{r}
U_{CN}^{i-k , i} $, the corresponding Hamiltonian at each site $i$ turns out to be
$H_i ^f = H_N ^i + \sum_{k=1}^{r}( H_{CN}^{i+k , i} +
H_{CN}^{i-k , i} ).$ In view of the generators of the quantum gates given above,
the Hamiltonian model of the QSCA reads,
\be
H_f =  \sum_{i\geq 0} \frac{1}{2}[ (\sigma_x^i + \sigma_z^i )+ 
\sum_{k=1}^{r}  ({\bf 1}^{i+k} - \sigma_z^{i+k}) (\sigma_x^i - {\bf 1}^i ) +
({\bf 1}^{i-k} - \sigma_z^{i-k}) (\sigma_x^i - {\bf 1}^i ) ].
\ee

\nd This can be interpreted as the Hamiltonian operator of a infinite
quantum spin
chain model  with free boundaries and interactions ranged over $r$ neighbors.
Obvious modifications such that QCSA with periodic boundary conditions,
or higher dimensions are in the present formalism appealing and worth
of further study.

{\it Quantum Fast Rule Theorem.} We turn now to study QSCA in terms of the so called Fast Rule Theorem(FRT)
\cite{frt} that provides analytic tools for the classification and the
study of types of dynamical evolution of classical SCA. In the limited
space available here we shall confine ourselves to the special case
of periodic particles and provide a quantum circuit of operators that
governs the time development of periodic qubit-particles, i.e the
quantum analogue  of periodic particles of classical CA theory\cite{cca,takh,bso}.

We recall that a particle is an integer multiple of $\ \ r+1$ consecutive
sites and it can be a collection of the so called {\it basic strings}
(BS), which are $r+1$ consecutive sites starting with a boxed site.
If a particle consists of a single BS, we call it a {\it simple particle}.

\nd Then a basic theorem states the following\cite{cca}:  Let a single particle
$OA^1 A^2 \cdots A^L O$ with $A^1\neq O , A^L \neq O$, consisting
of $L$ basic strings $A^1 , A^2 , \cdots , A^L $ (here $O$ denotes the null
BS consisting of $r+1$ zeros), and let $\oplus$ denotes
sitewise $\tt XOR$ among bits of the BS. Suppose that the BS's $A^1 , A^1\oplus A^2 ,
A^2 \oplus A^3 , \cdots ,A^{L-1}\oplus A^L , A^l $ , contain $l_0 , l_1 , \cdots , l_L $
$1$'s respectively. Then  if such a particle does not split or loose any BS's from its
right end at all times during its evolution, then at times $t=l_0
+l_1 +\dots l_m, \quad m\leq k \le L $, it becomes
$A^{m+1}\oplus (A^{m+2}A^{m+3}\cdots A^L A^0 A^1 \cdots A^m )$. Especially for 
$k=L$, i.e at time  $t=l_0 +l_1 +\dots l_L $ it returns to its initial state
$A^1 A^2 \cdots A^L $.

To illustrate the workings of the quantum FRT i.e  the quantum adaptation of the preceding theorem, let us consider
an initial particle of $L=3$ classical BS's $A^1 A^2 A^3$. Such a classical particle  by means
of the quantization map $\rho$  becomes the qubit word
$\ket{A^1 A^2 A^3 OOOO}$. Let us assume that the classical word labelling this state vector evolves
under the previously stated 
conditions so that the classical FRT applies and assures that
it has a periodic behaviour. Then the quantization map induces
into the quantum state vector the periodic evolution of its classical label word.
Namely the quantum state vector evolves periodically, and we seek to determine 
the evolution operator that implemends this periodic motion.
Assuming we have a sufficient number of null
quantum BS's from each side of
the initial particle state i.e $\ket{A^1 A^2 A^3 OOOO}$, let us consider the following two operators.
First, the operator ${\cal P}_0^I =\sum_{x\in \zz^{* r+1}}\proj{0^{r+1}}{x}$, which
projects each $r+1$-fold tensor product BS state vector onto the $r+1$-fold null state vector. With the upper index $I$ denotes the fact that the $(r+1)$-fold input vector is 
placed in the $I$-th position of a given chain of tensor product of state vector, and that
${\cal P}_0^I$ acts non trivial only in that $I$-th subspace. Second, let us define
the collective $\tt CN$ gate operator 
${\cal U}_{CN}^{IJ }\equiv U_{CN}^{I_{r+1} J_{r+1}}\cdots
 U_{CN}^{I_{2} J_{2}}U_{CN}^{I_{1} J_{1}}$,
with control BS placed in the $I$-th position  and
target BS placed in the $J$-th position. This collective $\tt CN$ gate is actually defined as a succesion of $(r+1)$ $\tt CN$ gates
for the corresponding qubits of the $(r+1)$-fold tensor product control and the target BS states vectors. 

In Fig. 2 we indicate graphically the sequence of operations that
tranform the initial state vector $\ket{A^1 A^2 A^3 OOOO}$ into the final state
$\ket{OOOOA^1  A^2  A^3 }$, indicating in this way its periodic $\it propagation $.
Each parallel wire in the figure represents an $(r+1)$-fold tensor product of qubits
and the operators ${\cal P}_0$ are indicated by boxed zeros, while operators 
${\cal U}_{CN}^{IJ }$
are indicated a double circled $\tt CN$ gate symbol. Explicitly the sequence of
operations of the quantum circuit in Fig. 2 reads:
\bea
\ket{A^1 A^2 A^3 OOOO} & \rightarrow &  \nn \\
\ket{OA^1 \oplus A^2  A^1 \oplus A^3  A^1  OOO} &=& 
 {\cal P}_0^1 {\cal U}_{CN}^{14}{\cal U}_{CN}^{13}{\cal U}_{CN}^{12}\ket{A^1 A^2 A^3 OOOO}\rightarrow \nn \\
\ket{OOA^2 \oplus A^3  A^2 A^1 \oplus A^2 OO}&=&{\cal P}_0^2 {\cal U}_{CN}^{25}{\cal U}_{CN}^{24}{\cal U}_{CN}^{23}\ket{OA^1 \oplus A^2  A^1 \oplus A^3  A^1  OOO }\rightarrow \nn \\
\ket{ OOOA^3  A^1 \oplus A^3  A^2 \oplus A^3  O}&=&{\cal P}_0^3 {\cal U}_{CN}^{36}{\cal U}_{CN}^{35}{\cal U}_{CN}^{34}\ket{ OOA^2 \oplus A^3  A^2 A^1 \oplus A^2 OO}\rightarrow \nn 
\\
\ket{ OOOOA^1  A^2 A^3}&=&{\cal P}_0^4 {\cal U}_{CN}^{47}{\cal U}_{CN}^{46}{\cal U}_{CN}^{45}\ket{ OOOA^3 A^1 \oplus A^3  A^2\oplus A^3 O}\;.
\eea

{\it Quantum parallelism.} The quantum description
of the cells of a CA in terms of qubits allows for having a
superposition $a\ket{0}+b\ket{1}$ at each cell. According to the standard interpretation
of Quantum Mechanics\cite{qm} this combination means that we have the state
$\ket{0}$($\ket{1}$)
with probability $|a|^2$($|b|^2$). This is the inherent
probabilistic character of QCA that entails two important
advantages of QCA over classical probabilistic (noisy) CA
(see e.g \cite{wolf}). First, there is an exponential
overhead in storing cell values in QCA over classical CA:
an $r$-radious probabilistic SCA needs at each discrete time
step  to store $N= 2^{2r+1}$ input words in order to 
process them later on and to update the current cell value. On the contrary
a QSCA may form and admit as input a superposition state vector $\ket{\psi}=\frac{1}{\sqrt{N}}
\sum_{x\in \zz_2^{* 2r+1}} \ket{x}\equiv \frac{1}{\sqrt{N}}
\sum_{x\in \zz_2^{* 2r+1}}
\ket{x_1 \cdots x_{2r+1}}$, which with
resources linear in $r$ i.e $2r+1$ qubits only  can constructs an
exponential in $r$ register of $2^{2r+1}$ equiprobable input states. 
Second, there is
an exponential overhead in the updating time of the cells. Once a superposition of all input
states has been prepared as the single state vector $\ket{\psi}$, then by acting
the linear evolution operator $\uca$ only once on it i.e $\uca\ket{\psi}=\frac{1}{\sqrt{N}}
\sum_{x\in \zz_2^{* 2r+1}} \uca\ket{x}=\frac{1}{\sqrt{N}}
\sum_{x\in \zz_2^{* 2r+1}} \ket{f(x)}$, we can update all the $N$ state vectors simultaneously.
This same exponential acceleration in storing space and processing time of quantum information has been known
in quantum computation, where it is utilized in a number of tasks such as fast quantum algorithms, information encoding, cryptography etc\cite{qcomp}. 

Harnessing the space-time resources made available by the 
quantum parallelism of QCSA in various applications is an open challenge. The research
program initiated here will require further studies of QSCA both as a new computational
machine and as an implementable physical system. Some of these questions however will be addressed
elsewhere\cite{forth}.

{\it Discussion}.
We close by discussing some prospects of our work. Towards a realization of QSCA as a physical
   process that could be implemented experimentally and so would
   potentially provide a novel quantum computational machine,
   we may construct an optical circuit made of elementary
   passive optical elements i.e light beam splitters and phase
   shifters that are assembled so that they implement the operator
   $U_f$ of a QSCA. For that purpose we may utilize the theorem of embedding
   the $SU(2)$ Lie group into $SU(N)$, which recently\cite{reck} has
   been used in the form of expressing every unitary matrix
   by a sequence of embedded $2 \times 2$ elementary unitary
   matrices of only two types : one matrix that realizes a
   beam splitter with
   transmission and reflection coefficients determined by the
   matrix elements and one that similarly realizes a phase
   shifter of a classical propagating light wave. By virtue of this
   embedding we may factorize the $U_f$ matrix of a $r=2$, say  QSCA
   into elementary matrices of beam splitters and phase shifters.
   Assuming a two-state encoding of the qubits of the QSCA cells
   (e.g taking the vertical and horizontal polarization states
   of a laser
   beam as $|0 >$  and $|1 >$), we may construct the quantum
   optical analog of the $U_f$ of the QSCA.

    Finally, we should mention a number of ramifications of
    our QSCA formalism that are currently also under
    investigation, namely: QSCA beyond the binary case
    ( this would involve higher dimensional representations
    of the discrete Heisenberg group ); differential versus matrix
    formulation of QSCA (i.e partial differential
    operators acting on a space of multivariable complex
    polynomials realizing the respective action of the
    $U_f$ matrix ); and also the case of {\it noisy QSCA}
    (i.e QSCA evolution that occurs in the presence of
    of quantum mechanical noise, exemplified by errors due
    to random bit-flipping and phase shifting in the qubits of the
    QSCA string ); this would require the description of
    QSCA in terms of trace-preserving operators instead of
    the unitary operators as in the noiseless case, and
    the respective quantization of the classical SCA to
    be carried out not
    by Hilbert-space state vectors but by employing the
    $\rho$-density
    operator formalism of Quantum Mechanics.

\vskip 0.5cm

%\pagebreak


\begin{thebibliography}{25}
\bibitem{wolf}
S. Wolfram, {\it Theory and Applications of Cellular Automata}
(World Scientific, Singapore 1986).
\bibitem{steig}
J. K. Park, K. Steiglitz and W. P. Thurston, Physica D {\bf 19},
423 (1986);\\
K. Steiglitz, I. Kamal and A. Watson, IEEE Trans. on Comp.
{\bf 37}, 138 (1988).\\
M. Jakubowski, K. Steiglitz and R. Squier, PhysComp96, (New England Complex Systems Institute, 1996), Eds. T. Toffoli et. al., p. 171.
\bibitem{frt}
T. S. Papatheodorou, M. J. Ablowitz and Y. G. Saridakis, Stud. Appl.
Math. {\bf 79}, 173 (1988).
\bibitem{cca}
%A. S. Fokas, E. P. Papadopoulou, Y. G. Saridakis and M. J. Ablowitz,
%Stud. Appl. Math. {\bf 81}, 153 (1989);
A. S. Fokas, E. P. Papadopoulou, Y. G. Saridakis, Physica D {\bf 41},
297 (1990);\\
A. S. Fokas, E. P. Papadopoulou, Y. G. Saridakis,
Phys. Lett. A
{\bf 147}, 369 (1990);\\
A. S. Fokas, E. P. Papadopoulou, Y. G. Saridakis,
Complex Syst.
{\bf 3}, 615 (1989).
\bibitem{takh}
M. J. Ablowitz, J. M. Keiser and L. A. Takhtajan, Phys. Rev. A. {\bf 44}, 6909 (1991).
\bibitem{bso}
M. Brushi, P. M. Santini and O. Ragnisco, Phys. Lett. A. {\bf 169}, 151 (1992).
%M. Brushi, P.M. Santini, Physica D, {\bf 70}, 185 (1994).
\bibitem{soliton}
F. Calogero and A. Degasperis, {\it Spectral Transform and Solitons}
(North-Holland, Amsterdam 1982).
%\bibitem{soli}
%M. J. Ablowitz ang H. Segur, {\it Solitons and Inverse
%Scattering Transform} Vol. 4 (SIAM, Philadelphia, PA 1981);
%A. C. Newell, {\it Solitons in Mathematics and Phyiscs} Vol. 45
% (SIAM, Philadelphia, PA 1985).
\bibitem{qm}
A. Peres, {\it Quantum Theory: Concepts and Methods}
( Kluwer, Dordrecht 1995).
\bibitem{qcomp}
A. K. Ekert and R. Jozsa, Rev. Mod. Phys. {\bf 68}, 733 (1996);\\
Eds. H.-K Lo, S. Popescu and T. Spiller, {\it Introduction to Quantum Computation and Information}, (World Scientific, Singapore 1998), and references therein.
\bibitem{bar}
A. Barenco, et. al Phys. Rev. A {\bf 52}, 3457 (1995).
\bibitem{forth}
D. Ellinas, E. P. Papadopoulou and Y. G. Saridakis, forthcoming.
\bibitem{reck}
M. Reck, A. Zeilinger et. al., Phys. Rev. Lett. {\bf 73}, 58 (1994).
\end{thebibliography}
\end{document}